\documentclass[%
 reprint,
superscriptaddress,
 aps,
 prl,
 floatfix
]{revtex4-2}

\usepackage[braket, qm]{qcircuit}
\usepackage{braket}
\usepackage{float}

\usepackage{dcolumn}
\usepackage{bm}
\usepackage{mathtools} 
\usepackage{color}  
\usepackage[T1]{fontenc}
 \usepackage[capitalize]{cleveref}
 
\definecolor{mscolor}{rgb}{0,0.5,0.5}

\definecolor{tgcolor}{rgb}{0.5,0,0.5}
\definecolor{cpcolor}{rgb}{0.4,0,0.8}
\definecolor{rkccolor}{rgb}{0.5,0.5,0}

\definecolor{lpcolor}{rgb}{0.3,0.3,0.3}

\usepackage{xcolor} 
\newcommand {\rsub}[1]{\textcolor{black}{#1}}

{}
\definecolor{phcolor}{rgb}{0.5,0,0.5}

\newcommand{\UWMPhysics}{Department of Physics, University of Wisconsin-Madison,  Madison, WI, 53706}
\newcommand{\UWMECE}{Department of Electrical \& Computer Engineering, University of Wisconsin-Madison,  Madison, WI, 53706}
\newcommand{\UCB}{Department of Physics, University of California Berkeley, Berkeley, CA, 94720}

\newcommand{\Infleqtion}{Infleqtion,  Madison, WI, 53703}

\begin{document}

 \title{Laser cooling and qubit measurements on a forbidden transition in neutral Cs atoms }

\author{J. Scott}
\affiliation{\UWMPhysics} 
\author{H. M. Lim}
\affiliation{\UWMPhysics} 
\author{U. Singla}
\affiliation{\UWMPhysics} 
\author{Q. Meece}
\affiliation{\UWMPhysics} 
\author{C. Fang}
\affiliation{\UWMECE}
\author{J. T. Choy}
\affiliation{\UWMECE}
\author{S. Kolkowitz}
\affiliation{\UCB} 
\author{T. M. Graham}
\affiliation{\UWMPhysics} 
\author{M. Saffman}
\affiliation{\UWMPhysics}
\affiliation{\Infleqtion}

\date{\today}

\begin{abstract}
We experimentally demonstrate background-free, hyperfine-level-selective measurements of individual Cs atoms by simultaneous cooling to $5.3~\mu\rm K$ and imaging on the  $6s_{1/2}\rightarrow 5d_{5/2}$ electric-quadrupole transition. We achieve hyperfine resolved detection with fidelity 0.9993(4) and atom retention of 0.9954(5), limited primarily by vacuum lifetime. Performing state measurements in a 3D cooling configuration enables repeated low loss measurements. A theoretical analysis of an extension of the demonstrated approach based on quenching of the excited state with an auxiliary field, identifies  parameters for hyperfine-resolved measurements with a projected fidelity of $\sim 0.9995 $ in $\sim 60~\mu\rm s$.
\end{abstract}

\maketitle


\noindent
\textcolor{blue}{{\it Introduction}} 
Fast, high-fidelity measurement of qubit states is a requirement for scalable quantum error correction. While great progress has been made towards fast and high-fidelity one- and two-qubit gates with neutral atom qubits \cite{Evered2023,Radnaev2025,Tsai2025,Peper2025,Muniz2025}, state measurements in qubit arrays with high fidelity and low loss currently require integration times of at least several ms, which is  3-4 orders of magnitude slower than gate times. While fast and high fidelity state measurements have been demonstrated by positioning atoms in optical cavities \cite{Bochmann2010,Volz2011,Deist2022,Grinkemeyer2025,BHu2025}, integrating atom arrays with cavities in a scalable fashion remains an outstanding challenge, and we consider here only free space approaches to measurement. Very fast measurement of single atoms in free space has also been demonstrated recently, but without resolving the internal states, and without retaining the atom after the measurement \cite{LSu2025}. 
An overview of free-space measurement results is shown in Fig.~\ref{fig.measurement}, highlighting the trade-off between measurement time and measurement fidelity. Thus, speeding up qubit measurements while retaining high fidelity and low atom loss is a critical requirement for improving the computational power of neutral atom quantum computers. 

In the context of quantum computing, detection of the presence or absence of an atom is useful for qubit register preparation, while detection of the internal atomic state is necessary for syndrome extraction within an error correction cycle. 
In this letter we demonstrate a new, background-free approach to measurement of qubits encoded in the hyperfine states of neutral Cs atoms.
Background-free imaging of atoms is possible by detecting fluorescence light at a wavelength different from that used for excitation. This was proposed in early work using transitions from metastable levels in alkaline earth atoms \cite{Loftus2002} and demonstrated in several experiments with clouds of alkali atoms \cite{Sheludko2008,Ohadi2009,BYang2012,McGilligan2020,LLI2024}, as well as with single trapped atoms \cite{Menon2024}, and with trapped ions \cite{Linke2012,Lindenfelser2017}. Previous demonstrations of background-free imaging with neutral atoms did not provide a state-selective measurement. 
 
\begin{figure}[!t]
\center
\includegraphics[width=.99\columnwidth]{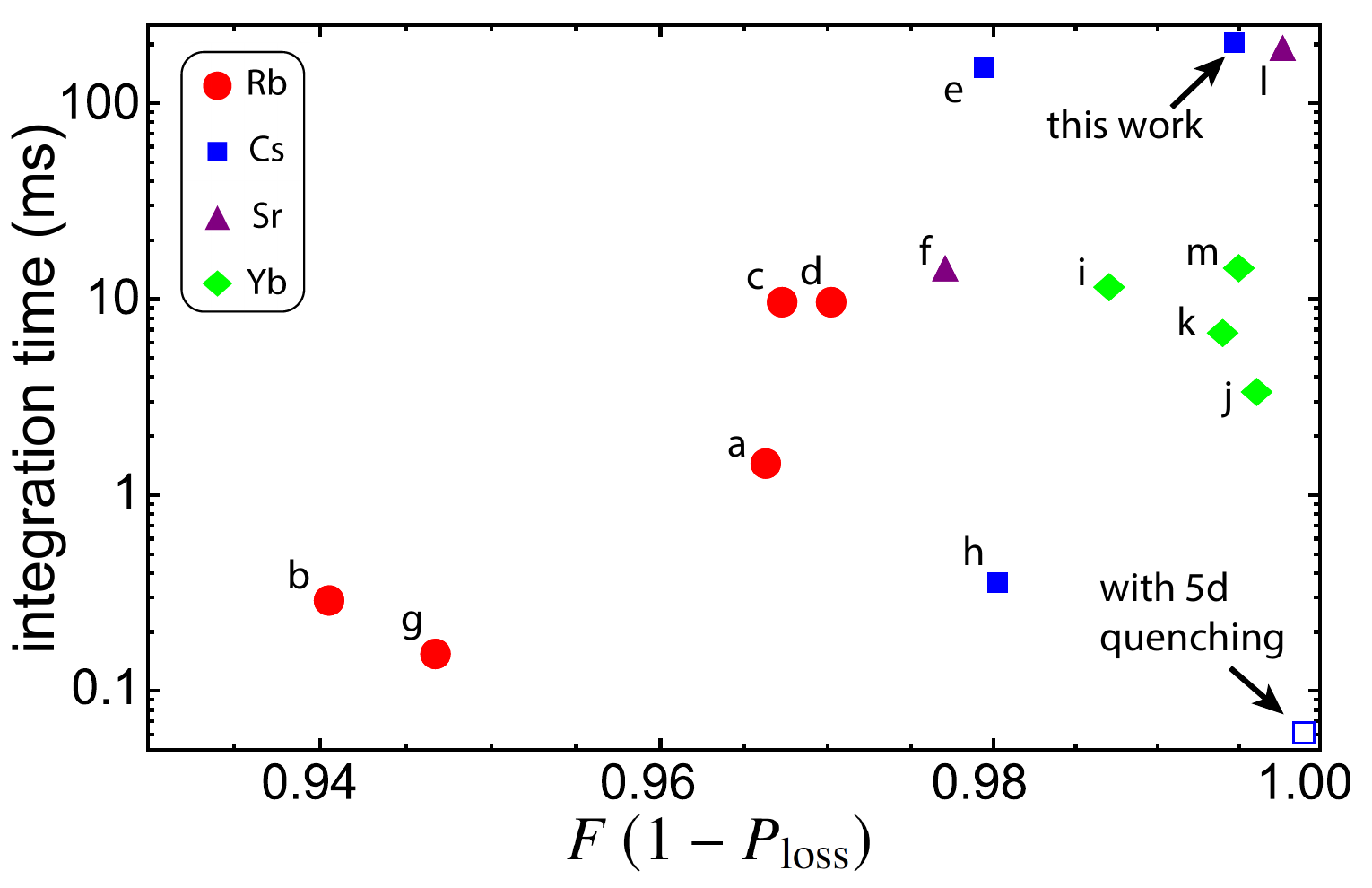}
\vspace{-.7cm}
\caption{\label{fig.measurement} Representative measurements of neutral atom qubit states in free space reported in the literature. The results are quantified in terms of the measurement time and the generalized fidelity given by the state detection fidelity $F$ times the atom retention probability $1-P_{\rm loss}$ . The experiments a,b,g,h  with the shortest integration times used single photon counting detectors and all others used cameras:  
a \cite{Fuhrmanek2011},
b \cite{Gibbons2011},
c \cite{Kwon2017},
d \cite{Martinez-Dorantes2017},
e \cite{TYWu2019},
f \cite{Covey2019a},
g \cite{Shea2020},
h \cite{Chow2023},
i \cite{Huie2023},
j \cite{Lis2023},
k \cite{Norcia2024},
l \cite{RTao2024},
m \cite{SMa2023}. The open square shows the predicted performance with excited state quenching.}
\end{figure}

Here we present a technique for background-free, and hyperfine-level-selective measurements in Cs atoms based on excitation of the electric-quadrupole (E2) transition from the ground $6s_{1/2}$ level to  $5d_{5/2}$. This transition is also of interest as the basis for an optical clock with Cs atoms \cite{Sharma2022}.  
Our demonstrated measurement approach is state selective while providing cooling along all three axes to prevent heating-induced trap loss. The state-selective character is due to the use of the $5d_{5/2}$ excited state, which has a large ratio of $y=\Delta_{\rm hf}/\Gamma$, where $\Delta_{\rm hf}$ is the excited state hyperfine splitting and $\Gamma$ is the radiative linewidth. The rate at which photons can be scattered while preserving the initial hyperfine state, relative to the rate of state changing Raman events scales as $y^2$, and this factor is $\sim 1000\times$ larger for scattering from $5d_{5/2}$ compared to the typical $6p_{3/2}$ level used for Cs atom imaging.  
 
 Our experimental approach yields some of the highest fidelity values reported for neutral atom qubits in free space, and to our knowledge the highest fidelity demonstrated with an alkali atom, but at the expense of long integration times ($\sim$200 ms) due to the low scattering rate from $5d_{5/2}$ and some excess technical noise detailed below. In order to circumvent the scattering rate limitation, we show theoretically that the long-lived $5d_{5/2}$ state can be quenched by stimulating the dipole-allowed transition from $5d_{5/2}$ to $6p_{3/2}$. We project that the scattering rate can thereby be increased by a factor of 50 or more with only minimal increase in unwanted ground state hyperfine changing transitions. Master equation simulations quantitatively verify the performance while taking full account of the three-dimensional nature of the driving fields and atomic sublevels.

\begin{figure}[!t]
\center
\includegraphics[width=.99\columnwidth]{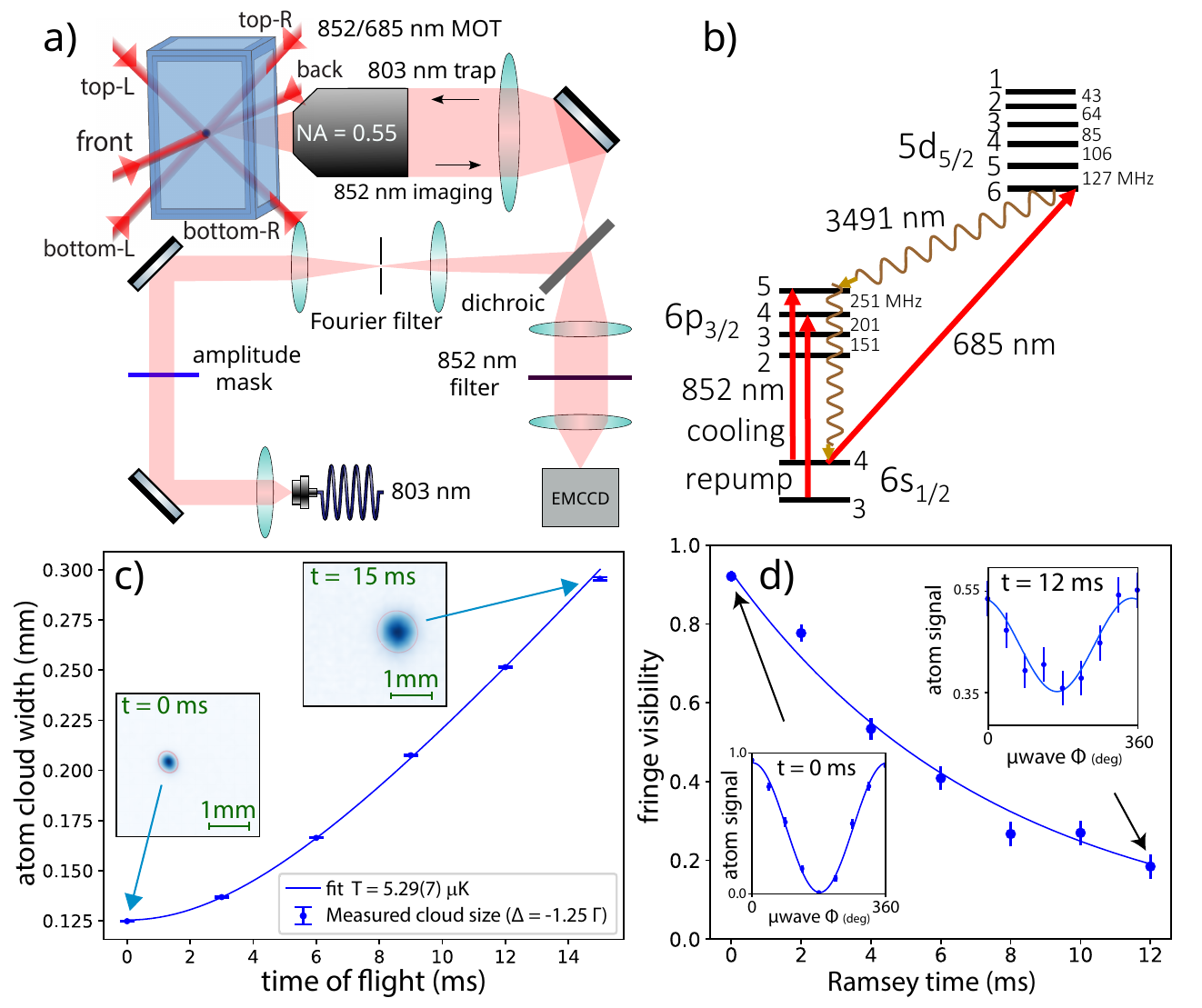}
\vspace{-.5cm}
\caption{\label{fig.setup} Laser cooling on the Cs 685 nm quadrupole transition. a) Experimental setup with magnetic quadrupole axis along the front-back beams and magnetic bias field along the axis of the objective lens, b) transitions used for first and  2$^{\rm nd}$ stage cooling, c) time of flight measurement of temperature of 2$^{\rm nd}$ stage MOT, d) Ramsey coherence of a single atom cooled with 685 nm light. \rsub{The fit functions in panels c) and d) are defined in the sections on MOT temperature and $T_2^*$ measurements in  \cite{SMScott2025}.}}
\end{figure}

 \noindent \textcolor{blue}{{\it Experimental approach}}
     The experimental apparatus, see Fig.~\ref{fig.setup}, involves a 2D magneto-optical trap (MOT) and a 3D MOT in a differentially pumped upper chamber. A near-resonant ``push'' beam  is used to transfer atoms from the 2D to the 3D MOT. We implement the MOTs using the Cs $6s_{1/2} - 6p_{3/2}$, 852 nm transition. The $6s_{1/2} - 5d_{5/2}$ quadrupole line 685 nm light is combined with the 3D MOT 852 nm beams using dichroic filters. Details concerning  beam parameters are given in  \cite{SMScott2025}.
     To trap single atoms, we use a 803 nm, bottle-beam trap implemented using the imaged hole array technique presented in  \cite{Huft2022}. In this work, a single hole is imaged into the cell as opposed to a large array. The trap depth was estimated to be $110~\mu\rm K$ for ground state atoms and the radial vibrational frequency was measured to be 10.6 kHz.   At 803 nm, the polarizabilities of $6s_{1/2}$ and $5d_{5/2},f=6$ are equal, allowing for a magic trapping condition as described in  \cite{Sharma2022}. An objective lens with numerical aperture  NA$=0.55$ is used to project the trap light into the glass vacuum cell, and to collect 852 nm fluorescence light for measurements.  The collected light is imaged onto an EMCCD camera. Narrow line filters with high transmission at 852 nm are used to remove stray trap and 685 nm light.
 
\noindent
\textcolor{blue}{ {\it Quadrupole cooling} }
After preparing a 3D MOT with the 852 nm cooling light using standard techniques, 2$^{\rm nd}$  stage cooling on the quadrupole transition was applied. The $5d_{5/2}$ state has a radiative linewidth of $\Gamma/2\pi=118~\rm kHz$ \cite{Pucher2020}, with essentially 100\% percent of the decay via the dipole allowed transition to $6p_{3/2}$ at 3491 nm. The Doppler temperature of this transition is  \rsub{$T_{\rm D}=\left(\frac{\hbar\Gamma}{2 k_{\rm B}} \right)\frac{k_{685}^2 + k_{3491}^2+k_{852}^2 }{2 k_{685}^2}= 2.39 ~\mu\rm K$. The recoil temperature of the 685 nm transition is $T_{\rm rec}=0.307~\mu\rm K$ so $T_{\rm D}/T_{\rm rec}=7.8$ whereas a quantum simulation with a simplified model \cite{Kirpichnikova2019} predicts a minimal temperature of 
$\sim 10 T_{\rm rec}.$ We defer a study of alternative configurations that may lead to sub-Doppler cooling on this transition to future work.} 
As shown in Fig.~\ref{fig.setup}c time-of-flight measurements following a 20 ms quadrupole cooling phase give an atom temperature of $5.29~\mu\rm K$. Simulations based on the theory developed in  \cite{Kirpichnikova2019} extended to account for the full hyperfine structure of the participating levels predict a temperature of about $4~\mu\rm K$ for the experimental parameters we used. 
Note that since the $5d_{5/2}, f=6$ level decays via $6p_{3/2},f=5$ to the $f=4$ ground level, and excitation of $5d_{5/2}, f\le 5$ is strongly suppressed, no repump light from $6s_{1/2},f=3$ is used for 2$^{\rm nd}$ stage cooling.  Additional details of the experimental approach are provided in \cite{SMScott2025}. 

\begin{figure*}[!t]
\center
\includegraphics[width=18.cm]{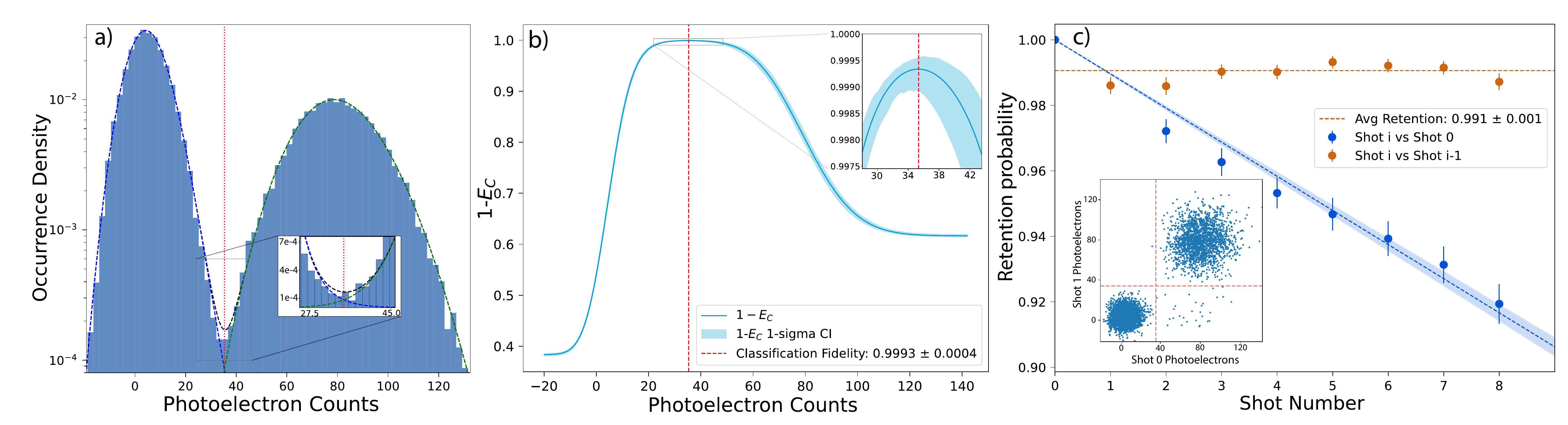}
\vspace{-.8cm}
\caption{\label{fig.state_measurement} Single atom state measurements with 685 nm 3D molasses. a) Histogram of photoelectron counts with 200 ms integration time, $\Delta=-2.54\Gamma$, $\sum_j I_j = 1.8~\rm W/cm^2$. The atom loading fraction is 41\%. b) State characterization error $E_c$ as a function of the count threshold \cite{Radnaev2025} c) Atom retention in the $f=4$ level after repeated measurements. The inset shows a two-measurement density plot with a few loss events \rsub{(lower right quadrant)} and no reloading events \rsub{(upper left quadrant)}. }
\end{figure*}

\begin{figure}[!t]
\center
\includegraphics[width=.95\columnwidth]{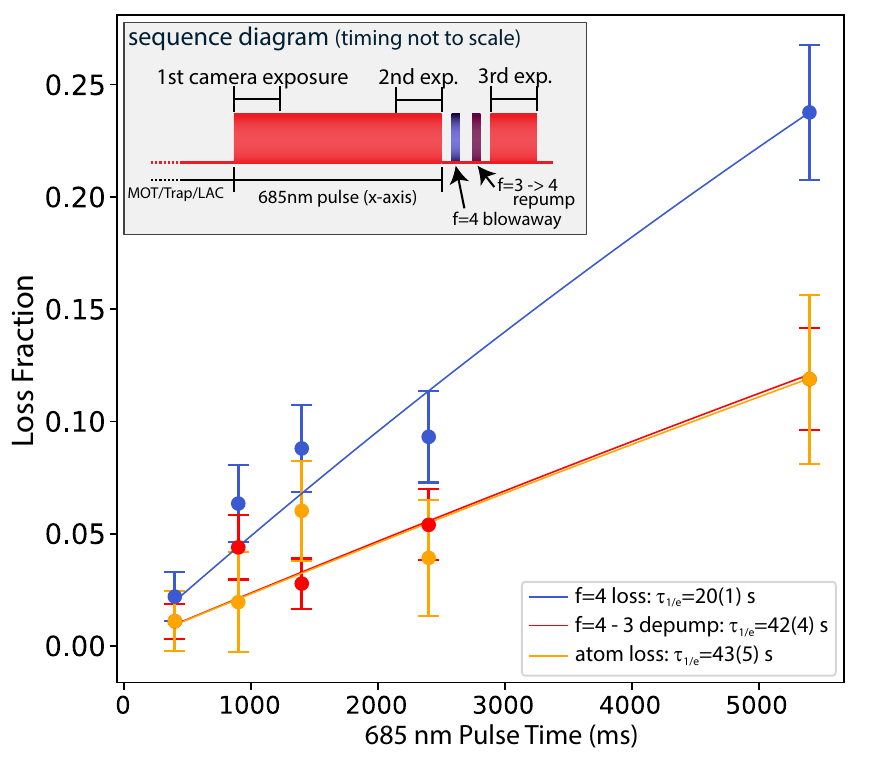}
\vspace{-.5cm}
\caption{\label{fig.depump} Measured atom loss and hyperfine depumping as a function of the duration of the quadrupole imaging pulse.  At the beginning and end of the pulse, the occupancy of $\ket{6s_{1/2}, f=4}$ is measured, \rsub{which is labeled as camera exposure}. To measure the hyperfine depumping fraction a resonant $\ket{6s_{1/2}, f=4} \rightarrow \ket{6p_{3/2}, f=5}$ ``blow-away'' beam removes all remaining $f=4$ atoms, and a short repump pulse moves all depumped atoms back into the $f=4$ level followed by a third imaging pulse. \rsub{See \cite{SMScott2025} for additional explanation.}}
\end{figure}

To perform cooling of single atoms we load into the 803 nm bottle beam trap and apply 3D cooling with the 685 nm light. 
The atom temperature was characterized by Ramsey coherence measurements. The atom is pumped into the $\ket{6s_{1/2},f=4,m_f=0}$ state with linearly polarized 895 nm light resonant with the $\ket{6s_{1/2},4,0} \leftrightarrow\ket{6p_{1/2},4,0}$ transition. Clock transitions between the $\ket{4,0} - \ket{3,0}$  ground states are then driven with a 9.2 GHz microwave field giving a $1/e$ Ramsey coherence time of 7.6(3) ms (see Fig.~\ref{fig.setup}d). The analysis in Ref.  \cite{Kuhr2005} is used to convert this into an upper bound on the atom temperature of $T_{\rm a}\le 5.4(2)~\mu\rm K$.  Since the optical pumping step prior to the Ramsey measurement heats the atom by up to several $\mu\rm K$ we estimate the temperature reached by 685 nm cooling in the bottle beam trap as being close to the Doppler limit.

\noindent  \textcolor{blue}{{\it Single atom state measurement}}
 To verify the state measurement performance a sequence of 10 images of a trapped atom is taken with no hyperfine repump light.   For the data shown in Fig.~\ref{fig.state_measurement} each imaging cycle consists of a 200 ms integration time at detuning $\Delta=-2.54 \Gamma$ and total intensity $I=\sum_j I_j= 1.8 I_{\rm sat}$, with $I_{\rm sat}=0.879 ~\rm W/cm^2$ the Zeeman and polarization averaged saturation intensity for the $\ket{6s_{1/2},f=4}-\ket{5d_{5/2},f''=6}$ transition (see  \cite{SMScott2025} for calibration and calculation details). After each measurement there is a 20 ms cooling phase, again with no hyperfine repumper,  at $\Delta=-2.8 \Gamma$. The magnetic fields are kept constant with a magnitude less than $10^{-6}~\rm T$. 
 The imaging parameters imply that 5,250 photons are scattered in 200 ms corresponding to a rate of $26,000~\rm s^{-1}$. With a calibrated total detection efficiency of $\eta=0.014$ \cite{SMScott2025} we expect an average photoelectron signal of 74 counts which agrees well with the observed separation of 75 counts between the bright and dark peaks in Fig.~\ref{fig.state_measurement}a).  The agreement is better than expected given the $\sim 10\%$ uncertainty in the optical intensity at the atom due to the use of narrow imaging beams and broadening of the $\ket{5d_{5/2},f=6}$ level from Zeeman and tensor shifts which broaden the observed lineshape to about 200 kHz.

 The observed background photoelectron count distribution in Fig.~\ref{fig.state_measurement}a) has a standard deviation of  $\sigma\simeq 7$ photoelectrons, while the bright distribution has $\sigma\simeq 15$ photoelectrons.
 Additional measurements with the 803 nm trap light off reveal that about 2/3 of the background noise is due to spectrally broad scattering from either the objective lens or the vacuum cell windows. With improved components and tighter spectral filtering around the 852 nm detection wavelength this noise contribution could be substantially reduced. Doing so would  reduce the required integration time to close to 100 ms, without loss of measurement fidelity. We discuss other strategies for faster measurement below.

The sequential images allow us to measure the atom retention fidelity by repetitively imaging the same atom.  Figure \ref{fig.state_measurement} illustrates the state measurement results. The classification fidelity of the atom state is ${\mathcal F}=0.9993(4)$, based on fitting to the observed bright and dark photoelectron distributions following the analysis in  \cite{Radnaev2025}. 
 From Fig.~\ref{fig.state_measurement}c), there is a probability of $P_4=0.991(1)$ of retaining the atom in the $\ket{6s_{1/2}, f=4}$ level after the 200 ms imaging pulse. There are two primary loss channels for the atoms in the trap. The first is depumping into the $f=3$ ground level caused by scattering into the hyperfine changing decay path $\ket{5d_{5/2}, f''<6} \rightarrow \ket{6p_{3/2},f'<5}\rightarrow \ket{6s_{1/2},f=3}$. The other loss mechanism is loss of atoms from the trap due to collisions with untrapped background atoms and molecules. Figure \ref{fig.depump} presents lifetime measurements of the atoms due to these two channels. The loss from $f=4$ is accounted for by both the depumping \rsub{lifetime} ($\tau_{f, 1/e}=42(4)~\rm  s$) and the trap lifetime ($\tau_{{\rm loss}, 1/e}=43(5) ~\rm s$), giving $\approx 0.46 \%$ loss of the atom from the trap  in 200 ms. These measurements indicate that the probability of atom retention can be significantly improved by shortening the measurement  time and increasing the vacuum lifetime.

\noindent \textcolor{blue}{{\it Measurement speedup by excited state quenching}}
A useful qubit measurement scheme should provide fast cycling between a single state and an excited state to generate scattered photons, with low probability of depumping into non-cycling or dark states. For qubit states defined by the hyperfine ground states in Cs or other alkali atoms, a common choice for state-selective cycling  is the D2 line between the $\ket{ns_{1/2},f=I+1/2}$ ground state and the $\ket{np_{3/2},f'=I+3/2}$ excited state, with $I$ the nuclear spin. The large GHz-scale separation between the hyperfine ground states prevents the same laser from driving  transitions out of $\ket{ns_{1/2},f=I-1/2}$, thus providing a fluorescence signal only when the atom is in the upper hyperfine level.  The much smaller  hyperfine splitting in the $np_{3/2}$ manifold $(\Delta_{\rm hf})$, typically a few 100 MHz in the heavy alkali atoms, leads to undesired Raman depumping into the lower ground hyperfine level which limits measurement fidelity \cite{Kwon2017,Martinez-Dorantes2017}.

In order to compare different measurement schemes we introduce a figure of merit for the Raman limited measurement fidelity of  $N_\Gamma=r_{\rm cycle}/r_{\rm Raman}$ where $r_{\rm cycle}$ is the cycling rate and $r_{\rm Raman}$ is the Raman rate. 
In the limit of low saturation and small detuning 
$r_\text{cycle} \propto |\Omega|^2/\Gamma$ and   $r_\text{Raman} \propto \Gamma |\Omega|^2/\Delta_\text{HF}^2$, where $\Omega$ is the excitation Rabi rate. Thus 
$N_\Gamma\propto y^2$ with $y=\Delta_{\rm hf}/\Gamma$ as discussed in the introduction.  Exact calculations of $N_\Gamma$ including angular momentum factors are provided in  \cite{SMScott2025}. For $^{87}$Rb atoms cycling is possible using the $5p_{3/2}$ or $4d_{5/2}$ excited states which have $N_\Gamma=38,200$ and $27,500$ 
respectively. For Cs atoms we can use $6p_{3/2}$ or $5d_{5/2}$ which have $N_\Gamma=39,200$ and $18.5\times 10^6$. The exceptionally large value of $N_\Gamma$ for the $5d_{5/2}$ state is due to the combination of relatively large $\Delta_{\rm hf}$ and very small $\Gamma_{5d}$ which arises because of the long wavelength of the $5d_{5/2}\rightarrow 6p_{3/2}$ decay path and the $\omega^3$ factor in Fermi's golden rule.  

While large $N_\Gamma$ is desirable for high fidelity state measurements, as we have demonstrated above, small $\Gamma_{5d}$ leads to slow measurements. We can retain a large $N_\Gamma$ while speeding up the measurement by quenching the excited state lifetime through stimulated emission with an auxiliary field on the $\ket{5d_{5/2},f''=6}\leftrightarrow \ket{6p_{3/2},f'=5}$ transition at 3491 nm.
For the purpose of quantifying the achievable measurement time we assume that scattering of 100 photons on the cycling transition is sufficient to measure the quantum state with an assignment error less than $2.5\times 10^{-4}$. The value 100 is somewhat arbitrary but is based on the following assumptions for detection efficiency using a qCMOS camera: double sided atom imaging \cite{Graham2022} with a numerical aperture of NA$=0.7$ ($\eta_{\rm NA}=2\times 0.14$ of the solid angle), optical transmission efficiency from atoms to camera of $\eta_{\rm optics}=0.80$, and camera detection quantum efficiency of $\eta_{\rm det}=0.50$. This gives a combined detection efficiency of $\eta=\eta_{\rm NA}\eta_{\rm optics}\eta_{\rm det}=0.11$ which implies that 100 scattered photons will give an average of 11 photoelectron counts. Allowing for Poisson distributed signal and noise and following the analysis in Fig. 5 of  \cite{Petrosyan2024} the signal limited measurement error is $<2.5\times 10^{-4}.$

Optimization of parameters to achieve fast and high fidelity measurements was done by numerical solution of  the Lindblad master equation for the system dynamics \cite{SMScott2025}.  The simulations involve all hyperfine levels of $6s_{1/2}$, $6p_{3/2}$, and $5d_{5/2}$ with a bias magnetic field of $10^{-5}~\rm T$. 
The simulated 685 nm beam configuration includes two pairs propagating perpendicular to the atom's quantization axis with linear polarization along the axis, and one pair propagating along the axis with circular $\sigma_\pm$ polarization. 
A pair of auxiliary $\sigma_+$ polarized beams propagating along the quantization axis are added to drive the $\ket{5d_{5/2},f''=6} \rightarrow \ket{6p_{3/2},f'=5}$ transition.

Numerical results are shown in Fig.~\ref{fig.Cs_dstate}. The behavior is governed by four free parameters: the Rabi frequencies $(\Omega_{E_1}, \Omega_{E2})$ and detunings $(\Delta_{E_1}, \Delta_{E_2})$  for the dipole and quadrupole fields.   In panels a) and b) the detunings are set to  $\Delta_{E_1} = 2\pi\times 0.6$ MHz, $\Delta_{E_2} = -2\pi \times 0.42$ MHz which minimizes infidelity for a measurement time of $\sim 60 \ \mu$s, thereby constraining two of the four parameters in the system. The Rabi frequencies, which remain as free parameters, are used to color-code the data. We see that an infidelity approaching $5\times 10^{-4}$ can be obtained in $60~\mu\rm s$ measurement time. This corresponds to a scattering rate of $1.67\times 10^6~s^{-1}$ which is $4.5\times$ faster than the radiative limit of $\Gamma_{5d}/2$ and $64\times$ faster than the scattering rate of $26,000~\rm s^{-1}$ observed in experiment. 

\begin{figure}[!t]
\center
\includegraphics[width=0.99\columnwidth]{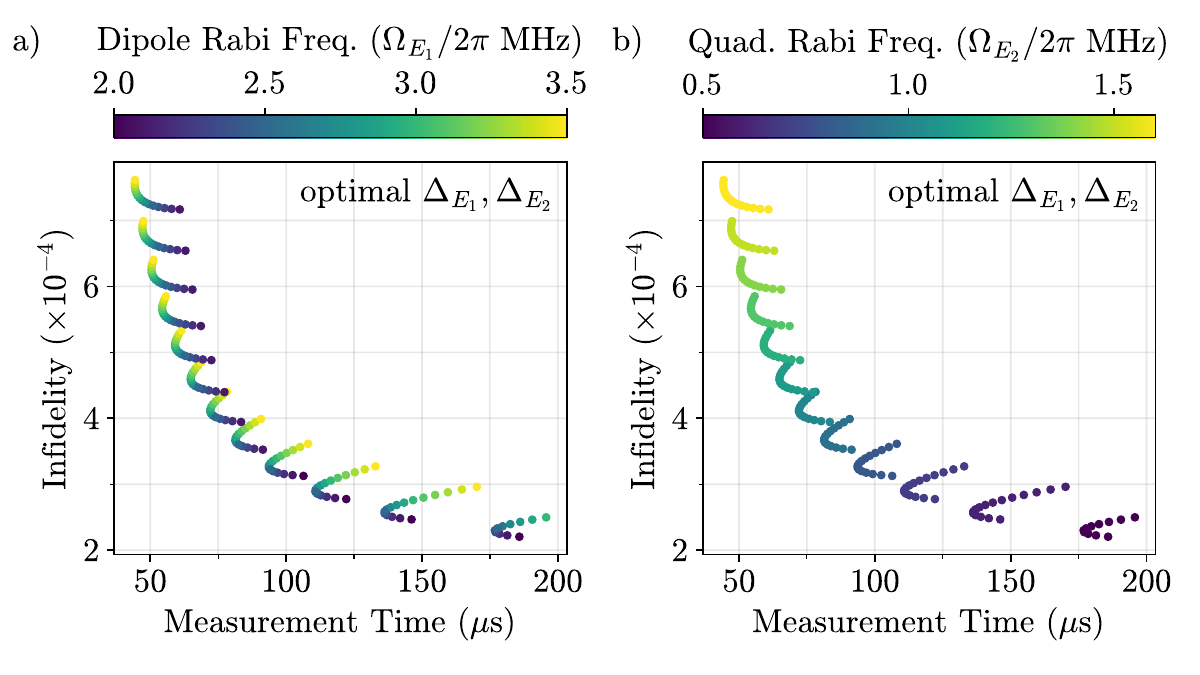}
\vspace{-.7cm}
\caption{\label{fig.Cs_dstate} Simulation of quadrupole imaging with auxiliary quenching field showing the dependence of measurement infidelity due to Raman events and the measurement time needed to scatter 100 photons. The underlying dataset is identical in both panels, with colors indicating $\Omega_{E1}$ in panel (a) and $\Omega_{E2}$ in panel (b).  The detunings were set to $\Delta_{E_1} = 2\pi\times 0.6$ MHz, $\Delta_{E_2} = -2\pi \times 0.42~{\rm MHz} = -3.56\, \Gamma_{5d}$. A minimum infidelity of $5.03\times 10^{-4}$ at a measurement time of $60~\mu\rm s$ is found for $\Omega_{E_1} = 2\pi\times 2.8$ MHz and $\Omega_{E_2} = 2\pi\times 1.2$ MHz. The dipole Rabi frequency of the quenching beam is defined  on the transition $\ket{5d_{5/2},f''=6,m_{f''}=1} - \ket{6p_{3/2},f'=5,m_{f'}=0}.$ The quadrupole Rabi frequency is defined on the transition $\ket{6s_{1/2},f=4,m_{f}=0} - \ket{5d_{5/2},f''=6,m_{f''}=\pm1}. $ 
 }
\end{figure}

\noindent \textcolor{blue}{{\it Conclusion}} 
In summary we have demonstrated high fidelity and low loss readout of a Cs atom hyperfine qubit while simultaneously cooling the 3D motion using background free imaging on a quadrupole transition. The demonstrated measurement time is relatively slow, limited by the small linewidth of the excited state and excess noise in the experimental apparatus. Based on numerical studies, we project that it is possible to reduce the measurement time to $60~\mu\rm s$ or less while retaining high fidelity measurement capability by quenching the excited state through stimulated emission with an auxiliary driving field. Combining the approaches demonstrated and analyzed here with a multiple atom repetition code as described in  \cite{Petrosyan2024}, as well as the use of machine learning enhanced image processing \cite{Phuttitarn2024} has the potential to reduce the measurement time to well under $10~\mu\rm s$.

\noindent \textcolor{blue}{{\it Acknowledgments}}
While completing this work we became aware of a  complementary demonstration  of Cs atom imaging using the same quadrupole transition \cite{Blodgett2025}. 
This work was supported by  
NSF  QuSeC-TAQS award 2326784,
the Wisconsin Alumni Research Foundation, 
Infleqtion, 
NSF Grant No. 2016136 for the QLCI Hybrid Quantum Architectures
and Networks, 
and the U.S. Department of Energy Office of Science National Quantum Information Science Research Centers as part of the Q-NEXT center.  The gold aperture used to create the optical trap was fabricated in the UW-Madison Wisconsin Center for Nanoscale Technology (wcnt.wisc.edu), which is partially supported by the Wisconsin Materials Research Science and Engineering Center (NSF DMR-2309000) and the University of Wisconsin-Madison.


\pagebreak
.

\newpage

\setcounter{page}{1}
\renewcommand{\theequation}{SM.\arabic{equation}}
\renewcommand{\thesection}{SM.\arabic{section}}
\setcounter{equation}{0}


{\bf Supplemental Material for:\\ Laser cooling and qubit measurements on a forbidden transition in neutral Cs atoms }


\date{\today}

The supplemental material provides details of supporting calculations and measurements.

\maketitle

\begin{widetext}

\tableofcontents

\section{Calculating $N_\Gamma$}
\label{sec.N_gamma}
$N_\Gamma$ is defined in the main text as the expected number of photons scattered in a readout cycle before a single depumping event occurs. In this section, we present an exact calculation of $N_\Gamma$, assuming the excitation light to be unpolarized. Under this assumption, the effective Rabi frequency for excitation from a state $\ket{f}$ to $\ket{f'}$ is defined as, 
\begin{align}
    \label{eq.effectiveRabi}
    \left|\Omega^{(i)}_{f'\leftarrow f}\right|^2 = \frac{1}{(2f+1)}\sum_{q, m_f} \left|\Omega^{(i)}_{f', m_f +q; f, m_f, q}\right|^2
\end{align}
where $q$ denotes the light polarization state  and $i \in \{1,2\}$ corresponds to the dipole and quadrupole Rabi frequencies, respectively. The light intensity is assumed to be equally distributed among all polarization states. The expressions in eqs. (\ref{eq.dipole_rabi},\ref{eq.quadrupole_rabi}), appropriately modified to account for changes in polarization, can be used to evaluate the effective Rabi frequency. This formulation enables us to express the scattering rate for a particular hyperfine transition as,
\begin{equation}
    r^{(i)}_{f'\leftarrow f}
    = \frac{\Gamma}{2} \frac{\frac{2|\Omega|^2}{\Gamma^2}}{1 + \frac{4\Delta^2}{\Gamma^2} + \frac{2|\Omega|^2}{\Gamma^2}} \bigg|_{\Omega = \Omega^{(i)}_{f'\leftarrow f}}.
\end{equation}
Finally, the branching ratio $b_{f'\rightarrow f}$ for spontaneous decay from $\ket{f'}$ to $\ket{f}$ is given by, 
\begin{equation}
    b_{f'\rightarrow f} = (2j'+1)(2f+1) \left|S^{j I f}_{f'1j'}\right|^2.
\end{equation}
\rsub{Here we have used the notation $S^{abc}_{def}=\left\{\begin{matrix}a&b&c\\d&e&f\end{matrix}\right\}$, where $\left\{...\right\}$ is a 6J symbol.}

With the necessary components in place, we can now model the rates of the cycling and hyperfine-changing Raman processes  \cite{Kwon2017}. When using the $6p_{3/2} \ (f' = 2, 3,4, 5)$ states for cycling, the rates of the processes that transfer population from $f_a = 4$ to $f_b = 3,4$ can be expressed as,
\begin{equation}
    R^{(1)}_{a\rightarrow b} = \sum_{f'} r^{(1)}_{f'\leftarrow f_a} b_{f'\rightarrow f_b}.
\end{equation}
On the other hand, when using the $5d_{5/2} \ (f'' = 6, 5, 4, 3, 2)$ states for cycling, where the population decays to $6p_{3/2} \ (f' = 2, 3, 4, 5)$ before returning to the hyperfine ground states, the rate of population transfer is given by
\begin{equation}
    R^{(2)}_{a\rightarrow b} = \sum_{f'} \sum_{f''} r^{(2)}_{f''\leftarrow  f_a} b_{f''\rightarrow  f'} b_{f'\rightarrow  f_b}.
\end{equation}
$N_\Gamma$ can then be calculated as the ratio of the cycling and Raman processes,
\begin{align}
    N^{(i)}_\Gamma = \frac{r^{(i)}_\text{cycle}}{r^{(i)}_\text{Raman}} = \frac{R^{(i)}_{4\rightarrow 4}}{R^{(i)}_{4\rightarrow 3}}.
\end{align}
In the limit of low saturation and small detuning, we obtain
\begin{align}
    N^{(1)}_\Gamma &= 3.92 \times 10^4\\
    N^{(2)}_\Gamma &= 1.85 \times 10^7,
\end{align}
verifying that the use of the $5d_{5/2}$ level  substantially reduces the Raman rate between ground hyperfine levels. An analogous calculation for $^{87}$Rb atoms gives $N^{(1)}_\Gamma = 3.82 \times 10^4$ for the $5p_{3/2}$ excited state and $N^{(2)}_\Gamma = 2.75 \times 10^4$ for imaging via the $4d_{5/2}$ state. 

\begin{figure}[!t]
\center
\includegraphics[width=.6\columnwidth]{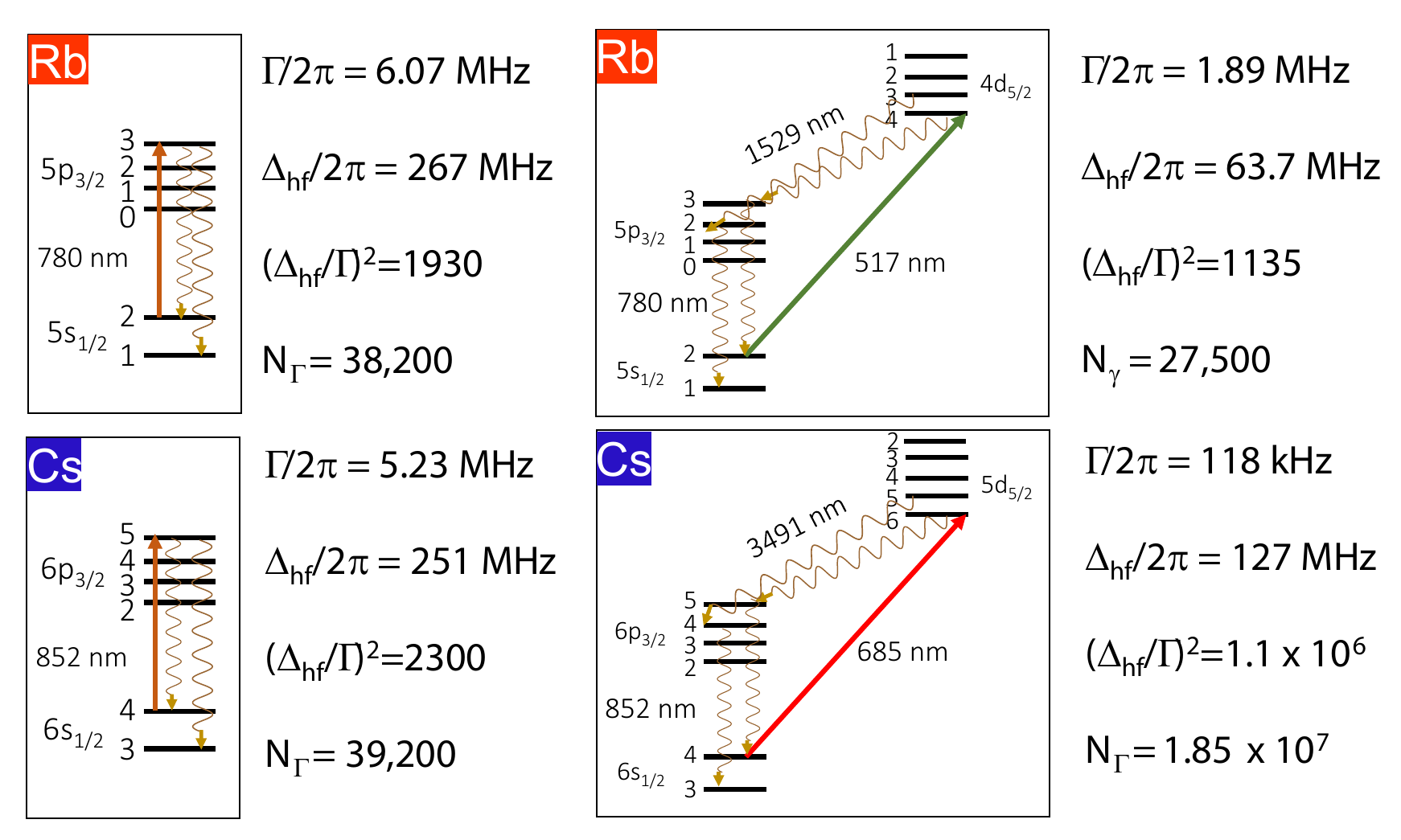}
\caption{\label{fig.Ngamma} Atom measurement parameters for the D2 and quadrupole transitions of $^{87}$Rb and Cs atoms.  }
\end{figure}

\section{Depumping rate analysis}
\label{sec.depump_rate}

Section \ref{sec.N_gamma} provides the analytical framework for evaluating the depumping rate as a function of the Rabi frequency and the detuning of unpolarized light. In order to compare with experimental measurements, we rewrite the scattering rate in a more convenient form using saturation intensity rather than Rabi frequency as,
\begin{equation}
    r^{(2)}_{f'', f} (I, \Delta)
    = \frac{\Gamma}{2} \frac{\frac{I}{I_\text{sat} (f'', f)}}{1 + \frac{4\Delta^2}{\Gamma^2} + \frac{I}{I_\text{sat} (f'', f)}}
    \label{eq.rscat}
\end{equation}
where $I_\text{sat} (f'', f)$ is the saturation intensity for the hyperfine transition from $f$ to $f''$. Using the effective quadrupole Rabi frequency $\left|\Omega^{(2)}_{f'', f}\right|$ between hyperfine states, averaged over Zeeman sublevels as defined in Eq. (\ref{eq.effectiveRabi}), the saturation intensity is given by
\begin{equation}
    I_\text{sat}(f'', f) = \frac{\epsilon_0 c \hbar^2 \Gamma^2}{4e^2k^2 \left|\Omega^{(2)}_{f'', f}\right|^2}.
\end{equation}
The saturation intensities $I_\text{sat}(f'', f{=}4)$, expressed in W/cm$^2$, are calculated to be 0.879, 1.484, 2.821, 6.529 and 22.852 for decreasing values of $f''$ from 6 to 2, respectively.

The depumping rate can then be evaluated as,

\begin{align}
    R_\text{depump} 
    &= R^{(2)}_{4\rightarrow 3} \\ 
    &= \sum_{f'} \sum_{f''} r^{(2)}_{f'', 4} (I, \Delta - \Delta_{\text{hf}, f''}) ~ b_{f'', f'} b_{f', 3}
\end{align}
where $\Delta$ denotes the laser detuning relative to the $f''=6$ hyperfine level and $\Delta_{\text{hf}, f''}$ is the hyperfine splitting between $f''$ and $f''=6$. 
At the experimental values of $ I = 1.8$ W/cm$^2$ and $\Delta = -2\pi\times 400$ kHz, the calculated depumping rate is 0.04 s$^{-1}$, indicating a depumping probability of 0.008 in a 200 ms duration measurement. This is about 60\% larger than the observed depumping rate in Fig.~\ref{fig.depump} of the main text. We attribute the discrepancy to uncertainty of the actual beam intensity at the location of the atom due to alignment uncertainties of the cooling beams with small mm sized waists. 

\rsub{
\section{Determination of state measurement fidelity and atom loss}}

\rsub{ 
Following \cite{Radnaev2025} we analyze the recorded photoelectron distribution from the EMCCD camera as 
\begin{align}
    P(x) &= A_{D}P_{D}(x) + (1-A_{D})P_{B}(x)\\
    P_{D}(x) &= \frac{1}{2K\sigma_D} \exp\left(\frac{1}{2K^2} - \left(\frac{x-\mu_D}{K\sigma_D}\right)\right)\mathrm{erfc}\left(-\frac{(x-\mu_D)/\sigma_D-1/K}{\sqrt{2}}\right)\\
    P_{B}(x) &= \frac{1}{\sqrt{2\pi}\sigma_B}\exp\left(-\frac{(x-\mu_B)^2}{2\sigma_B^2}\right)\left[1 + \mathrm{erf}\left(a\frac{x-\mu_B}{\sqrt{2}\sigma_B}\right)\right]
\end{align}
Here, $P(x)$ is the total photo-electron counts,  $P_{\rm D}(x)$ is the dark distribution weighted  by the fraction of dark counts $A_{\rm D}$, and $P_{\rm B}(x)$ is the bright distribution weighted by the bright fraction $1-A_{\rm D}$. 
The dark distribution is parameterized by $\mu_D$, $\sigma_D$, and $K$, which describe the location, scale, and shape of the dark distribution, respectively. The bright distribution is parameterized by $\mu_B$, $\sigma_B$, and $a$, which describe the location, scale, and shape of the bright distribution, respectively; $\mathrm{erf}(x)$ and $\mathrm{erfc}(x)$ are the error function and the complementary error function.}

\rsub{ 
We obtained the following parameter values after fitting the  distribution shown in Fig.~\ref{fig.state_measurement}b):  $A_D = 0.616(2),~ \mu_D = 1791.3(2),~ \sigma_D = 6.42(8), ~K=0.59(4),~ \mu_B = 1858.5(9),~ \sigma_B = 19.4(7),~ a = 1.2(2)$. 
Once the distribution is fit to the data, the optimal discrimination threshold $x_*$ is  determined as the photo-electron count that minimizes the sum of classification errors $E_{C}(x_*)$, which is
\begin{align}
E_{C}(x_*) = A_{D}\int_{x_*}^{\infty} dx\,  P_{D}(x) + (1-A_D)\int_{-\infty}^{x_*}dx\,  P_{B}(x).
\end{align}
}
\rsub{This analysis resulted in the optimal count threshold and state discrimination fidelity reported in Fig. \ref{fig.state_measurement}. }

\rsub{Figure \ref{fig.depump}  presents results for the rate of depmping to $f=3$ and the rate of atom loss out of the trap.  The measurement sequence involves the same sequence, repeated, for all curves. The $f=4$ loss is determined by the fraction of loaded atoms not seen in the second shot. The $f=4\rightarrow 3$ depumping loss is determined by the fraction of loaded atoms not seen in the second shot but seen in the third shot, since they have been protected from blow-away in the $f=3$ state then repumped to $f=4$. The trap loss fraction is determined by the fraction of loaded atoms not seen in both the second and third shots, meaning they are no longer present in the trap. The 685 nm pulse used is the same pulse as for the 685 nm imaging parameters. Thus, a ``shot" simply means that the EMCCD camera is triggered to acquire an image during that duration.}

\rsub{The measured depumping and atom loss time constants are both more than 40 s.   We have not separately measured the atom lifetime in the absence of the 685 nm pulse. Since the 685 nm light cools the atoms thereby reducing any heating due to intensity noise on the trap laser the  lifetime without the 685 nm light is likely to be shorter, and limited by the background vacuum conditions.  }\\

\section{Lindblad master equation simulation}
The dynamical population of the  levels involved in the state measurement  can be evaluated by solving the Lindblad master equation,
\begin{equation}
    \frac{\partial \rho}{\partial t} = - \frac{i}{\hbar} [\mathcal{H}, \rho] + \text{decay terms}
\end{equation}
where $\rho$ is the density operator and $\mathcal{H}$ is the system Hamiltonian.

The total Hamiltonian $\mathcal{H}$ for the quadrupole transition state measurement is 
\begin{equation}
\mathcal{H} = \mathcal{H}_0 + \mathcal{H}_{E_1} + \sum_{n = 1}^6\mathcal{H}^{(n)}_{E_2}.
\end{equation}
 $\mathcal{H}_0$ is a diagonal operator that includes the hyperfine energy of each level and the corresponding linear Zeeman shifts. Tensor shifts have not been included and are small for the dark trap used in the experiment.  The dipole allowed transitions  $\ket{5d_{5/2},f''} - \ket{6p_{3/2},f'}$ with $f''-1\le f'\le f''+1$ and $2\le f''\le 6$ driven by the 3491 nm auxiliary field  are accounted for by $\mathcal{H}_{E_1}$ and 
 $\mathcal{H}^{(n)}_{E_2}$ describes the quadrupole transition arising from the $n^\text{th}$ 685 nm laser beam.

To represent elements of the density operator, we assign integer labels to the hyperfine-Zeeman levels in order of increasing energy. The decay terms, hereafter denoted by $\mathcal{D}$, are added to the Lindblad equations as follows,
\begin{align}
    \mathcal{D}_{jj} &= - \sum_{i=1}^{j-1} \Gamma_{j\rightarrow i} \rho_{jj} + \sum_{i=j+1}^{N} \Gamma_{i\rightarrow j} \rho_{ii} \\
    \mathcal{D}_{jk} &= -\frac{1}{2} \left( \sum_{i=1}^{j-1} \Gamma_{j\rightarrow i} + \sum_{i=1}^{k-1} \Gamma_{k \rightarrow i }\right) \rho_{jk}
\end{align}
where $\Gamma_{i \rightarrow j}$ denotes the spontaneous decay rate from level $i$ to $j$.

We exclude the $5d_{5/2} \ (f=1)$ state from our simulations as it does not participate in the quadrupole excitations under consideration. This results in a system comprising $N=93$ levels.  We exploit the hermiticity of $\rho$ to write equations only for elements $\rho_{ij}$ where $i\ge j$, thus reducing the number of equations to about half. The off-diagonal terms of the density operator, also known as coherences, rotate with the applied fields, which warrants the substitution $\rho_{ij} = \tilde \rho_{ij} e^{-i\omega_{ij}t}$ where $\omega_{ij}$ is $\omega_1$ for states $(i, j) \equiv (5d_{5/2}, 6p_{3/2})$, $\omega_2$ for $(i, j) \equiv (5d_{5/2}, 6s_{1/2} (f=4))$, and $\omega_1-\omega_2$ for $(i, j) \equiv (6p_{3/2}, 6s_{1/2} (f=4))$. States that are not connected by lasers are assumed to have no coherence, and hence not considered.  We ignore any fast rotating terms, invoking the  rotating wave approximation. Spontaneous emission from $5d_{5/2}$ to $6s_{1/2}$ is ignored because it is much slower than Rabi frequencies and other decay rates. Finally, we express the equations in terms of Rabi frequencies defined as,
\begin{align}
    \label{eq.dipole_rabi}
    \Omega^{(1)}_{f', m_f'; f, m_f} &= \frac{d_{f', m_f'; f, m_f} \mathcal{E}^*_1}{\hbar} \\
    \label{eq.quadrupole_rabi}
    \Omega^{(2)}_{f', m_f'; f, m_f, n} &= \frac{Q^{(n)}_{f', m_f'; f, m_f} \mathcal{E}^*_2 k_2}{\hbar}.
\end{align}
Here $d$ and $Q$ are dipole and quadrupole matrix elements that are defined below. All six 685 nm beams are modeled to have identical $\ket{4, 0} \rightarrow \ket{6', \pm 1'}$ Rabi frequencies $\Omega^{(2)}_{6, \pm1; 4, 0, n}$, where the sign of $m_f'$ is determined by the polarization-dependent selection rules. 

We numerically integrate a system of 1899 coupled differential equations to track the population dynamics of the hyperfine Zeeman sublevels during the measurement process. All population is assumed to be in the $\ket{6s_{1/2},f=4,m_f=0}$  state at the start of the simulation. The number of scattered photons is calculated as $\Gamma_{6p}\times P$, where $P$ is the time integrated population in all sublevels of $6p_{3/2}$ during the measurement. The Raman error probability is calculated as the population in the $\ket{6s_{1/2},f=3}$ level at the end of the measurement.

\subsection{Dipole matrix elements}

Electric dipole (E1) transitions are driven by an electric field ${\bf E}_1 = \frac{\mathcal{E}^*_1}{2}e^{i({\bf k}_1 .{\bf r} - \omega_1t)} \rsub{\boldsymbol{\epsilon}_1} +  \frac{\mathcal{E}_1}{2}e^{-i({\bf k}_1 .{\bf r} - \omega_1t)}  \rsub{\boldsymbol{\epsilon}^*_1}$ with polarization $\rsub{\boldsymbol{\epsilon}_1}$, wavevector ${\bf k}_1$ and angular frequency $\omega_1$, appear in the Hamiltonian as, 
\begin{align}
    \mathcal{H}_{E1} &=  -{\bf d} \cdot {\bf E}_1 \\&\approx \frac{e}{2} \left( \mathcal{E}^*_1 e^{-i\omega_1t}{\bf r}\cdot \rsub{\boldsymbol{\epsilon}_1} + \mathcal{E}_1 e^{i\omega_1t}{\bf r}\cdot \rsub{\boldsymbol{\epsilon}^*_1} \right).   
\end{align}
The polarization of the dipole beam is chosen to be $\sigma_+$. Using the electric dipole matrix elements given by
\begin{equation}
     d_{f', m_{f'}; f, m_{f}} = \bra{\gamma', f', m_{f'}}  -e{\bf r}\cdot \boldsymbol{\epsilon}_1\ket{\gamma, f, m_f}
\end{equation}
we can write the Hamiltonian matrix elements as 
\begin{equation}
\label{eq.dipole_H}
    \bra{\gamma', f', m_{f'}}\mathcal{H}_{E_1}\ket{\gamma, f, m_f} = \\- \frac{1}{2} \left( \mathcal{E}^*_1 e^{-i\omega_1t} d_{f', m_{f'}; f, m_{f}} + \mathcal{E}_1 e^{i\omega_1t} d_{f, m_{f}; f', m_{f'}}^* \right).
\end{equation}

\subsection{Quadrupole matrix elements}
\label{sec.quadrupole_element}

 In this section, we evaluate the quadrupole matrix elements for a transition driven by an electric field $\boldsymbol{E}^{(n)}_2 = \frac{\mathcal{E}^*_2}{2}e^{i({\bf k}_2^{(n)} \cdot{\bf r} - \omega_2 t)}  \rsub{\boldsymbol{\epsilon}^{(n)}_2} + \frac{\mathcal{E}_2}{2}e^{-i({\bf k}_2^{(n)} \cdot{\bf r} - \omega_2 t)}  \rsub{\boldsymbol{\epsilon}^{(n)*}_2} $ with wavevector ${\bf k}_2^{(n)}$, angular frequency $\omega_2$ and polarization vectors $\rsub{\boldsymbol{\epsilon}^{(n)}_2} $. Since there are six quadrupole beams, each with a distinct wavevector and polarization, we denote the beams with superscripts $(n)$.

The Hamiltonian for electric quadrupole transitions is
\begin{equation}
    \mathcal{H}^{(n)}_{E2} = i \frac{e}{2} \bigg[\mathcal{E}^*_2 e^{-i\omega_2 t}(\rsub{\boldsymbol{\epsilon}^{(n)}_2} \cdot {\bf r}) ({\bf k}_2^{(n)} \cdot {\bf r}) \\
    - \mathcal{E}_2e^{i\omega_2 t}(\rsub{\boldsymbol{\epsilon}^{(n)*}_2} \cdot {\bf r}) ({\bf k}_2^{(n)} \cdot {\bf r})\bigg].
\end{equation}
We define the quadrupole matrix element between states $\ket{\gamma', f',m_{f'}}$ and $\ket{\gamma, f, m_f}$ as
\begin{equation}
        Q^{(n)}_{f', m_{f'}; f, m_{f}} = \\ \bra{\gamma ', f', m_{f'}} 
    -ie(\rsub{\boldsymbol{\epsilon}^{(n)}_2}\cdot {\bf r}) (\rsub{{\bf k}^{(n)}_2}\cdot {\bf r}) \ket{\gamma, f, m_f}
\end{equation}
where $\gamma, \gamma'$ represent any additional quantum numbers required to specify the states. The elements of the interaction can then be written as,
\begin{equation}
\label{eq.quadrupole_H}
\bra{\gamma', f', m_{f'}}\mathcal{H}^{(n)}_{E_2}\ket{\gamma, f, m_f} = \\- \frac{k_2}{2} \bigg( \mathcal{E}^*_2 e^{-i\omega_2 t} Q^{(n)}_{f', m_{f'}, f, m_{f}} + \mathcal{E}_2 e^{i\omega_2 t} Q^{(n)*}_{f, m_{f}, f', m_{f'}} \bigg).
\end{equation}

In order to evaluate the quadrupole matrix element, we express the polarization and the wavevector in spherical co-ordinates as $\rsub{{\boldsymbol{\epsilon}}^{(n)}_2} = \sum_{\nu} \epsilon^{(n)}_\nu {\bf e}^*_\nu$ and ${\bf k}_2^{(n)} = k_2 \sum_{\mu} k_\mu^{(n)} {\bf e}^*_\mu$, respectively. The matrix element then takes the form,
\begin{equation}
    Q^{(n)}_{f', m_{f'}; f, m_{f}} = \\ 
    - i e\bra{\gamma', f', m_{f'}} 
    \left(r^2 \sum_{q=-2}^2 c_{-q}^{(n)} C_{2,q} \right) \ket{\gamma, f, m_f}
\end{equation}
where $c_q^{(n)} = (-1)^q \sqrt{2/3} \sum_{\mu \nu} C^{2q}_{1\mu1\nu} k_\mu^{(n)} \epsilon_\nu^{(n)}$ and $C_{k,q} = \sqrt{\frac{4\pi}{2k+1}} Y_{k,q}$. Here, $C_{j_1 m_1  j_2 m_2}^{j_3 m_3}$ are  Clebsch-Gordon coefficients and $Y_{k, q}$ are spherical harmonics.

The quadrupole matrix elements for all six beams involve evaluating terms of the form $\braket{\gamma', f', m_{f'}| ~r^2 C_{2,q}~| \gamma, f, m_f}$, which can be expressed using the Wigner-Eckart theorem as,
\begin{equation}
    \bra{\gamma', f', m_{f'}} r^2 C_{2,q} \ket{\gamma, f, m_f} =
    \frac{C^{f' m_{f'}}_{f m_f  2 q}}{\sqrt{2f'+1}} \bra{\gamma', f'} |r^2 C_{2} | \ket{\gamma, f}.
\end{equation}
The last term in the above expression is the reduced matrix element, which can be calculated as,
\begin{eqnarray}
    \bra{\gamma', f'} |r^2 C_{2} | \ket{\gamma, f} &=&
    (-1)^{I + j + j' + s + l' + f} \sqrt{2f+1} \sqrt{2f'+1} \sqrt{2j+1}\nonumber \\
    &\times& \sqrt{2j'+1} \sqrt{2l+1} \  C^{l' 0}_{l0 20}  S^{j I f}_{f' 2 j'}  S^{l s j}_{j' 2 l'} \nonumber\\ &\times& \int_0^{\infty}dr\, r^2 R^*_{n' l'} \;r^2 \; R_{nl} 
\end{eqnarray}
where $R_{nl}$ are radial wavefunctions and  $S^{abc}_{def}=\begin{Bmatrix}
        a & b & c \\
        d & e & f
    \end{Bmatrix}$ is the Wigner 6j symbol.

\subsection{Simulation results}

To determine the exact readout fidelities and measurement times under different dipole and quadrupole field Rabi frequencies and detunings, we solved the Lindblad master equations governing the system numerically.  The simulations include all hyperfine levels of $6s_{1/2}$, $6p_{3/2}$ and $5d_{5/2}$ with Zeeman splitting corresponding to 0.1 G of magnetic field. The 685 nm beam configuration includes two pairs propagating perpendicular to the atom's quantization axis with linear polarization along the axis, and one pair propagating along the axis with circular ($\sigma_\pm$) polarization. 

In order to provide a comparison case we first simulated imaging on the D2 line with no quadrupole excitation. For this simulation we used the usual three pairs of $\sigma_+/\sigma_-$ counterpropagating beams at 852 nm. Representative results assuming the atom needs to scatter 100 photons are shown in Fig.   \ref{fig.Cs_pstate}. Here, panels (a) and (b) present the same data, colored according to Rabi frequency and detuning respectively, while panel (c) plots the measurement time and infidelity against Rabi frequency for $\Delta = - \Gamma/2$. We find that for a measurement time of $\sim 60 \ \mu$s and $\Delta = - \Gamma /2$, the minimum achievable infidelity in this scheme is $\sim 4 \times10^{-3}$. This is about an order of magnitude worse than the result shown in Fig.~\ref{fig.Cs_dstate} in the main text.

\begin{figure*}[!t]
\center
\includegraphics[width=0.96\columnwidth]{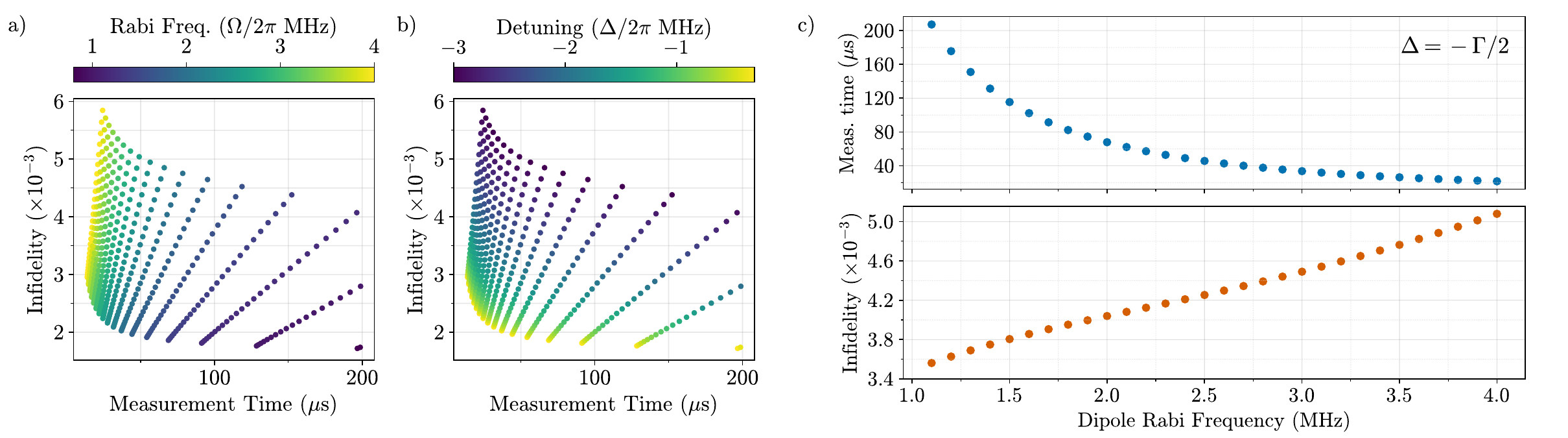}
\caption{\label{fig.Cs_pstate} Cycling via the $6p_{3/2}$ state for qubit readout. The atom is assumed to scatter 100 photons over the course of the measurement. Two key parameters are varied in the Lindblad simulation of this readout scheme: the Rabi frequency and the detuning from the zero-field $6s_{1/2}, f=4 \leftrightarrow 6p_{3/2}, f'=5$ resonance.  Panels $\boldsymbol{a}$ and $\boldsymbol{b}$ present identical data from parameter scans, colored by Rabi frequency and detuning, respectively. Each data point reflects a unique combination of these two values. In panel $\boldsymbol{c}$, the detuning is fixed at $-\Gamma/2$, and the effect of varying the Rabi frequency on both the measurement time and readout error is illustrated.}
\end{figure*}

Next we  simulated imaging by cycling through the $5d_{5/2}$ state, following the same procedure as for the $6p_{3/2}$ case. The results are presented in Fig.~\ref{fig.Cs_dstate_no_dipole}. In the presence of a 0.1 G magnetic field, the lowest-energy Zeeman state of the $5d_{5/2}, f''=6$ level is shifted approximately 420 kHz below its zero-field energy. The 685 nm quadrupole driving laser is red-detuned from this state in order to prevent atom heating during measurement. For the data presented in panels (a) and (b), the detuning is measured relative to the zero-field $6s_{1/2}, f=4 \leftrightarrow 5d_{5/2}, f''=6$ resonance. In contrast, panel (c) assumes a fixed detuning of $\Delta = -\Gamma/2$ relative to the lowest-energy Zeeman state. Our results indicate that measurement errors below $10^{-3}$ can be achieved -- surpassing the fidelity realized using using the $6p_{3/2}$ cycling scheme. However, this comes at the cost of longer measurement times, typically on the order of milliseconds. 
Depumping to the ground $f=3$ level in this scheme arises in two ways. 1)  Off-resonant quadrupole excitation of $5d_{5/2},f''<6$ followed by spontaneous decay to $6p_{3/2}, f'<5$ which can decay to $6s_{1/2}, f=3$. 
2) Sequential off-resonant stimulated transitions: An off-resonant excitation to $5d_{5/2}, f'' < 6$ may be followed by another off-resonant stimulated transition to the $6p_{3/2}, f'<5$ states, facilitated by the presence of the dipole field employed for enhancing the readout speed.

As we show in the main text Fig.~\ref{fig.Cs_dstate} including the quenching field in the simulation provides a similar fidelity to that shown in Fig.~\ref{fig.Cs_dstate_no_dipole} but with a $10\times$ faster measurement time. Note that for the full simulations the quenching field was defined to be a pair of $\sigma_+$ counterpropagating fields for cancellation of momentum kicks from stimulated emission.  In an implementation the beams could be detuned from each other by a few tens of kHz to prevent the formation of a standing wave.

\rsub{ For the quenching beam, a Rabi frequency of $\Omega / 2\pi= u$ MHz defined on the $\ket{5d_{5/2},f''=6,m_{f''}=1} \leftrightarrow \ket{6p_{3/2},f'=5,m_{f'}=0}$ transition is achieved with a laser beam intensity of $0.162 \times u^2$ mW/cm$^2$. At the reported optimal value of $u = 2.8$, the required intensity is 1.27 mW/cm$^2$, or $20\ \mu$W for a beam with $1~\rm mm$ waist. This low power requirement can be readily achieved with mid-infrared semiconductor laser sources. Furthermore, standard optical windows fabricated from fused silica glass have high transmission at the required  wavelength of approximately $3.49~\mu\rm m$.  }

\begin{figure*}[t]
\center
\includegraphics[width=0.96\columnwidth]{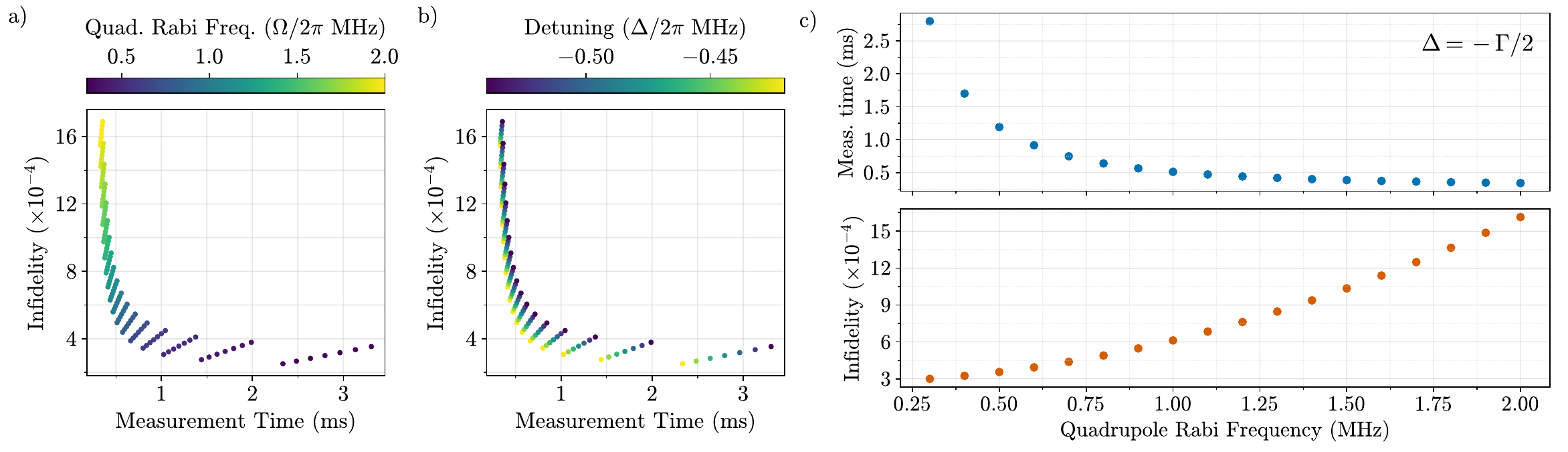}
\caption{\label{fig.Cs_dstate_no_dipole} Cycling via the $5d_{5/2}$ state for qubit readout with no dipole quenching field. Two key parameters are varied in the Lindblad simulation of this readout scheme: the Rabi frequency and the detuning from the zero-field $6s_{1/2}, f=4 \leftrightarrow 5d_{5/2}, f'=6$ resonance.  Panels a)  and b) present identical data from parameter scans, colored by Rabi frequency and detuning, respectively. Each data point reflects a unique combination of these two values. In panel c) the detuning is fixed at $-\Gamma/2$, and the effect of varying the Rabi frequency on both the measurement time and readout error is illustrated.}
\end{figure*}

\section{MOT  temperature measurement}
\label{sec.FreeSpaceQuad}

We measure the MOT temperature after free-space cooling on the quadrupole line using time-of-flight measurements. The first step is formation of a standard 852 nm MOT.  The 852 nm beam waists $(1/e^2$ intensity radius) were  3.2 mm  for the 4 diagonal top/bottom beams, and 3.7 mm for the front and back beams. 
 We then  perform a 20 ms quadrupole cooling phase involving ramping the magnetic field gradient down by a factor of 4 to $0.015~\rm T/m$  and turning on the 685 nm beams. 
The beams \rsub{at both 852 and 685 nm wavelengths} have standard $\sigma_+/\sigma_-$ polarizations for each counterpropagating beam pair \rsub{and the 685 nm beams had the same polarization state as the corresponding 852 nm beams.} A 1 kHz frequency shift is added for each (685 nm and 852 nm) counter propagating pair to prevent standing waves.
An optimal cooling temperature is found at $\Delta = -1.25 \Gamma_{5d}$, based on optimization of parameters for minimizing cloud expansion rate during time of flight.

The glass cell used in this experiment is antireflection coated but not for 685 nm which results in large, and incidence angle dependent reflection losses. The beam sizes, measured transmission values, and peak intensity on axis  inside the vacuum cell are given in Table  \ref{tab:cellchar}. 
The peak intensity was calculated from 
$$
I_{\rm peak}= \frac{2 P}{\pi w^2} T
$$
with $P$ the external beam power, $T$ the transmission from the table, and $w$ the beam waist. In order to account for small variations in the alignment of each beam to the location of the trapped atom while ensuring equal intensities,  the power in each beam was adjusted such that the atom blow away time for facing beam pairs was equal.  
The total peak intensity of the 685 nm light across all 6 beams is $1.80~{\rm W/cm^2}= 2.05 I_{\rm sat}$ where $I_{\rm sat}$ is the average saturation intensity for the $\ket{6s_{1/2},f=4} - \ket{5d_{5/2},f=6}$ transition.

\begin{table*}[!t]
 \caption{Cell window transmission data for the 685 nm cooling and imaging beams.  The transmission was determined by measuring the external  transmitted power through facing windows and assuming they have the same  transmission. The transmission column is the transmission of one window and the peak intensity is the estimated value at the atoms. The front and back beams had an incidence angle of 5 deg. and the top and bottom beams had an incidence angle of 50. deg. with respect to the window normal. The counterpropagating beam pairs were front/back, top-left/bottom-right, and top-right/bottom-left. }
    \label{tab:cellchar}
\centering
    \begin{tabular}{|l|c|c|c|}
    \hline
         Beam & Waist (mm) & Transmission (\%)& Peak Intensity $\rm (W/cm^2)$ \\ 
         \hline
        front &1.1 & 51  & 0.10 \\
        back &1.1 & 56  & 0.10 \\
        \hline
        top-left&0.58  & 92  & 0.23 \\
        bottom-right &0.58 & 85 & 0.47 \\
        \hline
        top-right &0.58 & 87  & 0.58 \\
        bottom-left &0.58 & 91 & 0.31\\
        
        \hline
        Total & &   & 1.80  \\
\hline
\end{tabular}
\end{table*}

The rate of expansion of the size of the atom cloud is used to calculate the temperature of the atoms using 
\rsub{ 
\begin{equation}
\label{eq.CloudSizevsAtomTemp}
    \sigma_{r}(t)=\sigma_{r}(0)\left(1+\frac{k_BT_{\rm a} t^2}{M\sigma_r(0)^2}\right)^{1/2}
\end{equation}}
where \rsub{$\sigma_{r}(t)$ is the root mean square width  of the  density distribution at time $t$}, $k_B$ is the Boltzmann constant, $M$ is the atomic mass, and $T_{\rm a}$ is the atom temperature. The data in  Fig.~\ref{fig.setup}c) in the main text  gives $T_{\rm a}=5.29(7)~\mu\rm K$ which is higher than the Doppler temperature of $2.39~\mu\rm K$.  
We leave further investigation of the atom temperature dependence on polarization, detuning, and intensity and comparison with theory \cite{Kirpichnikova2019} for future work.

\section*{Single atom trapping and fabrication of the intensity mask}

The Cs atom was trapped in a bottle  beam optical trap with 803 nm light. A single bottle beam trap was implemented using the method described in \cite{Huft2022}. 
A single gold aperture on an anti-reflection-coated glass substrate is imaged into the cell, as opposed to a large array. The aperture has a diameter of $200~\mu\mathrm{m}$ was fabricated using photolithography, deposition of a titanium adhesion layer and $100~\mathrm{nm}$ of gold (which gives near-zero transmittance at near-infrared wavelengths), liftoff, and oxygen plasma ashing.

\section{Measurement of $T_2^*$}
\label{sec.t2*}

We use the Ramsey method to measure $T_2^*$, the dephasing time between the clock states $\ket{f=3, m=0}$ and $\ket{f=4, m=0}$ within the ground state manifold of $6s_{1/2}$. After the optical trap is loaded with a single Cs atom, the atom's state is initialized to $\ket{6s_{1/2}, f=4,m=0}$ by optical pumping. A bias field of approximately $2.6\times 10^{-4}~\rm T$ is applied to lift the degeneracy of the Zeeman states. Then, two coherent resonant $\pi/2$ microwave pulses interleaved by a free evolution time are applied. The Rabi frequency of the microwave pulse is 11 kHz. 
The phase offset $\phi$ of the second $\pi/2$ microwave pulse is varied and contrast of the Ramsey fringes is read out by detecting the population  in the $f=4$ level.  The single atom measurement is repeated 200 times for each phase offset $\phi$.  

For each free evolution time, the contrast of the Ramsey fringes is obtained by fitting the measured atom population $n$ as a function of the phase offset $\phi$ according to 
\begin{align}
    n(\phi) =(1/2)[1+ A_t \cos(\phi+\phi_t)]
\end{align}
where $A_t$ is the fringe contrast  and $\phi_t$ is the phase accumulated during the free evolution time. 
The experiment is repeated for different free evolution times between 0 - 10 ms, and the contrast of the Ramsey fringes $A_t$ at different free evolution times is fit to
\begin{align}
    A_t= A_0 e^{-t/T_2^*}
\end{align}
to obtain the coherence time $T_2^*$. \rsub{We note that several dephasing mechanisms contribute to $T_2^*$ including  background magnetic fields, atom motion at finite temperature, and intensity noise on the trap laser \cite{Kuhr2005,Rosenfeld2011,Saffman2011}. Although the Ramsey fringe decay due to atomic motion alone is not exponential, contributions from Gaussian distributed magnetic and intensity noise are. We see in Fig. \ref{fig.setup}d) in the main text that the assumption of an overall exponential decay fits the data reasonably well and we have therefore used an exponential fit function.}  
As reported in the main text Fig.~\ref{fig.setup} we measured the coherence of a single trapped atom after quadrupole cooling to be $T_2^*=7.6(3)~\rm ms$ which implies a temperature of at most $5.4~\mu\rm K$ \rsub{using Eq. (28) in \cite{Kuhr2005}}.

\section{EMCCD Camera Characterization}
\label{sec.camera}

To calibrate the photon detection efficiency of the collection optics and the  camera, we image single atoms using the 852 nm transition for 70 ms exposure time, with $\sum_j I_j=2.5 \langle I_{\rm sat}\rangle$ where the Zeeman and polarization averaged saturation intensity for the Cs $\ket{6s_{1/2},f=4} - \ket{6p_{3/2},f=5}$ transition is $\langle I_{\rm sat}\rangle =2.7~\rm mW/cm^2$. The detuning was set to $\Delta=-5.64 \Gamma_{6p_{3/2}}$ where $\Gamma_{6p_{3/2}}=2\pi\times 5.23~\rm MHz$. With these parameters and Eq. (\ref{eq.rscat}) we calculate 22,000 scattered photons in a 70 ms pulse. 

The imaging train consisted of a NA$=0.55$ objective lens ($\eta_{\rm NA}=0.082$), an optical train including lenses, beamsplitters and filters with an estimated transmission efficiency of $\eta_{\rm optics}=0.8$, and an EMCCD camera with a manufacturer specified quantum efficiency of $\eta_{\rm det}=0.55$ at 852 nm. The EMCCD camera has a measured electron multiplying  gain of 178 and a conversion factor of 5.8 photoelectrons (after the electron multiplying gain) per output digital count value. A total of 100 EMCCD pixels are used to capture the photon signal. After analog  $2\times 2$ binning on the EMCCD the digital signal from 25 pixels is sent to the computer for summation to determine the state measurement.

We estimate a combined detection efficiency of $\eta=\eta_{\rm NA}\eta_{\rm optics}\eta_{\rm det}=0.036$ which implies that we should detect $22,000 \times 0.036 \simeq 790$ photoelectrons. 
We observe a separation between the dark and bright histogram peaks of $\sim 300$ photoelectrons. This suggests that there is  additional loss in either the optical train or the camera sensitivity at the level of 62\% such that the actual combined detection efficiency is $\eta=0.014$. This value of $\eta$ was used to estimate the actual photon scattering rate in the experiment. 

\end{widetext}

\bibliography{rydberg,optics,qc_refs,atomic,saffman_refs}

\end{document}